\documentclass[a4paper]{article}

\usepackage{cite}
\usepackage{graphicx}
\usepackage{latexsym}

\usepackage[all,cmtip]{xy}
\PassOptionsToPackage{hyphens}{url}\usepackage{hyperref}
\usepackage{epsfig,endnotes}

\usepackage{amsmath}
\usepackage{amsthm}
\usepackage{amssymb}
\usepackage{amsmath}
\usepackage{bm}
\usepackage{algpseudocode,amssymb,amsfonts,color,xspace}
\usepackage{pdfsync}
\usepackage{threeparttable}
\usepackage{color}
\usepackage[ruled, linesnumbered,lined,boxed,commentsnumbered]{algorithm2e}
\usepackage{url}

\usepackage[T1]{fontenc}
\usepackage[utf8]{inputenc}
\usepackage{authblk}

\usepackage{multirow}
\usepackage{threeparttable}
\usepackage[format=plain, 
justification=raggedright,singlelinecheck=false]{caption}

\usepackage{pgfplotstable}
\pgfplotstableset{
  empty cells with={---},
  every head row/.style={before row=\toprule,after row=\midrule},
  every last row/.style={after row=\bottomrule}
}
\pgfplotsset{compat=1.9}

\usepackage{array,booktabs}

\usepackage{enumerate}

\newcommand{\GF}[1]{\ensuremath{\mathbb{F}_{#1}}\xspace}

\newcommand{\TGF}[1]{\ensuremath{\mathbb{\widetilde{F}}_{#1}}\xspace}

\newcommand{\RR}[1]{\ensuremath{\mathbb{R}_{#1}}\xspace}

\newcommand{\xor}{\ensuremath{\texttt{XOR}\xspace}}
\newcommand{\pshufb}{\texttt{PSHUFB}\xspace}
\newcommand{\vpshufb}{\texttt{VPSHUFB}\xspace}
\newcommand{\pclmul}{\texttt{PCLMULQDQ}\xspace}

\newcommand{\lchbtfy}{\ensuremath{\texttt{Butterfly}\xspace}}
\newcommand{\ilchbtfy}{\ensuremath{\texttt{iButterfly}\xspace}}

\newcommand{\addfft}{\ensuremath{\texttt{addFFT}\xspace}}
\newcommand{\iaddfft}{\ensuremath{\texttt{iaddFFT}\xspace}}

\newcommand{\basiscvt}{\ensuremath{\texttt{BasisCvt}\xspace}}
\newcommand{\ibasiscvt}{\ensuremath{\texttt{iBasisCvt}\xspace}}

\newcommand{\enc}{\ensuremath{\texttt{Encode}\xspace}}
\newcommand{\dec}{\ensuremath{\texttt{Decode}\xspace}}

\newcommand{\novelpoly}{\emph{novelpoly}\xspace}

\newtheorem{theorem}{Theorem}[section]

\newtheorem{proposition}{Proposition}

\theoremstyle{definition}
\newtheorem{definition}[theorem]{Definition}

\newif\ifsubmission
\submissionfalse

\begin{document}

\title{Multiplying boolean Polynomials
with Frobenius Partitions in Additive Fast Fourier Transform
}

\ifsubmission
\author{}
\else

\author[1,3]{Ming-Shing~Chen}
\author[1]{Chen-Mou~Cheng}
\author[1,2]{Po-Chun~Kuo}
\author[2]{Wen-Ding Li}
\author[2,3]{Bo-Yin~Yang}
\affil[1]{Department of Electrical Engineering, National Taiwan University, Taiwan\\
\texttt{ \{mschen,doug,kbj\}@crypto.tw}
}
\affil[2]{Institute of Information Science, Academia Sinica, Taiwan\\
\texttt{ \{thekev,by\}@crypto.tw}
}
\affil[3]{Research Center of Information Technology and Innovation, Academia Sinica, Taiwan}

\fi

\maketitle

\begin{abstract}

We show a new algorithm and its implementation for multiplying bit-polynomials of large degrees.
The algorithm is based on evaluating polynomials at
a specific set comprising a natural set for evaluation with additive FFT
and a high order element under Frobenius map of $\GF{2}$.
With the high order element, we can derive more values of the polynomials
under Frobenius map. 
Besides, we also adapt the additive FFT to efficiently evaluate polynomials
at the set with an encoding process.

For the implementation, we reorder the computations in the additive FFT 
for reducing the number of memory writes and hiding the latency for reads.
The algebraic operations, including field multiplication, bit-matrix transpose, and bit-matrix multiplication,
are implemented with efficient SIMD instructions.
As a result, we effect a software of best known efficiency, shown in our experiments.

\end{abstract}



\section{Introduction}
\label{sec:introduction}

Multiplication for
long bit-polynomials in the ring $\GF{2}[x]$, where $\GF{2}$
is the finite field(or Galois field, GF) of two elements,
is a fundamental 
problem in computer science.
%
%
The operation is so basic that even modern CPUs
dedicates a hardware instruction 
for carryless multiplication of polynomials for a fixed size in modern CPUs.

To the best of our knowledge, all currently
fast algorithms for multiplication of long
bit-polynomials are based on
a fast Fourier transform (FFT) algorithm.
The FFT efficiently evaluates polynomials at subgroups in the underlying field,
and the multiplication of polynomials are performed
by evaluating polynomials at particular points, multiplying the
evaluated values from two inputs, and interpolating the values
back to a polynomial with inverse FFT algorithm.



\subsection{The FFTs Applied to Multiply Bit-polynomials}

Two categories of FFTs had been applied to multiply bit-polynomials.

The first category is ``multiplicative'' FFTs, evaluating 
polynomials at multiplicative subgroups formed by roots of unity.
For evaluating at $n$ points over fields of characteristic of 2(or binary fields),
unfortunately, a multiplicative subgroup of arbitrary size $n$ does not exist.
The construction of the desired subgroups becomes primary task and 
might induce an extra burden for the kind of FFTs.
For example, the Sch\"onhage \cite{schonhage1977} FFT evaluates polynomials
at points formed by a ``virtual'' root of unity with the order $n = 3^m$.
Harvey, van der Hoeven, and Lecerf \cite{DBLP:conf/issac/HarveyHL16}
presented another DFT(Cooley-Tuckey FFT) working on the specific field of $\GF{2^{60}}$,
allowing abundant multiplicative subgroups since $2^{60}-1$ has many small factors (smooth).

The other category is the additive FFTs which works on points forming an 
\emph{additive} subgroup.
For binary fields, the size of additive subgroups are $2^i$ for $i \in \mathbb{N}$ 
and exactly fits for divide-and-conquer FFTs.
The kind of FFTs had been developed by Cantor~\cite{Cantor:1989}, 
Gao and Mateer~\cite{gm-afft}, and Lin, Chung, and Han~\cite{lch-afft}
with the best known bit complexity $O(n \log n \log( \log n ) )$.
We will detail the additive FFT in sec.~\ref{sec:addfft}.

\subsection{Previous Approaches for the Multiplication}

\subsubsection{The practical multiplications for bit-polynomials}
On implementing the multiplication of bit-polynomials on modern computers,
since the computer works with instructions on machine words instead on a single bit,
the software usually works on a structure of multiple bits(e.g., 
a binary field of $m$-bits, denoted as $\GF{2^m}$) for higher efficiency.
Hence, while analyzing the complexity of algorithms,
the algebraic complexity model is more suitable than the bit complexity model.
In this model, from Harvey \emph{et al.}\cite{HHL17:polymul}, 
the best complexity for multiplying polynomials with degree $n$ 
is $O(n \log n )$ \emph{field multiplications} and $O(n \log n \log( \log n ))$ field additions
by Cantor and Kaltofen\cite{Cantor91onfast}.

In this paper, we discuss algorithms with the best known complexity and
supporting only the \emph{practical} length of polynomials, namely $n < 2^{64}$ bits.
The restriction is caused from working on a dedicated field instead of arbitrary fields.


\subsubsection{Kronecker Substitution of Coefficients of Polynomials}

Most previous works for multiplying bit-polynomials based on
the Kronecker substitution(KS)\cite{Gathen:2013:MCA:2512973}.
We denote the bit-polynomials of degree $< n$ as $\GF{2}[x]_{<n}$.
For computing $ A \cdot B \rightarrow C \in \GF{2}[x]_{<n}$ with KS,
we partition the $A$ and $B$ into $2n/m$ blocks of size $(m/2)$-bits, i.e.,
let $A = \sum_{i=0}^{(2\cdot n/m)-1} \hat{a}_i x^{i \cdot (m/2)}$ where $ \hat{a}_i \in \GF{2}[x]_{<m/2}$.
We then translate the $ \hat{a}_i$ as a field element in $\GF{2^{m}}$ and $\hat{A} \in \GF{2^{m}}[y]$ 
such that $A = \hat{A}(x^{m/2})$. We can then perform a standard polynomial multiplication with FFTs
over $\GF{2^{m}}[y]$. Note here we have to split the polynomials to blocks of size $(m/2)$
for preventing ``overflow''.

For multiplicative FFT implementations with KS, 
Brent \emph{et al.}\cite{gf2x} implemented mainly the
Sch\"onhage \cite{schonhage1977} algorithm in the library \texttt{gf2x}.
Harvey, van der Hoeven, and Lecerf \cite{DBLP:conf/issac/HarveyHL16} 
presented multiplication
using the DFT over the field $\GF{2^{60}}$, which 
size of elements closes to a machine word and
size of the field allows abundant multiplicative subgroups.
For additive FFT implementations,
Chen \emph{et al.}\cite{DBLP:journals/corr/abs-1708-09746}
presented a multiplication based on the
additive FFT over the fields of Cantor basis\cite{Cantor:1989}\cite{gm-afft}.
They utilized the subfield structure of the multipliers in the FFT and
further reduced the time taken for field multiplications.

\subsubsection{Frobenius Partitions of the Evaluated Points}
In 2017, van der Hoeven \emph{et al.}\cite{VLL2017:carrylesspolymul}
presented a new multiplier of two times improvement over their KS implementation\cite{DBLP:conf/issac/HarveyHL16}.
Instead of partitioning the polynomials in $\GF{2}[x]_{<n}$ into blocks, 
they directly translate the binary coefficients
into field elements of $\GF{2^m}$(specifically, $\GF{2^{60}}$ for $m=60$)
and performed the FFT at a special set $\Sigma_{\omega}$ of size $n/m$ in $\GF{2^m}$
instead of a larger set $\Omega_n$ of size $n$ points.
They showed the $\Sigma_{\omega}$ can construct $\Omega_n$ as well as
the corresponding evaluated values 
under the Frobenius map of $\GF{2}$, i.e., the square in binary fields.
By evaluating at only $|\Sigma_{\omega}| = n/m$ points in the DFT over $\GF{2^{60}}[x]$, 
they can thus accelerate the multiplication.
For multiplying polynomials in $\GF{2}[x]_{<n}$ and based field $\GF{2^m}$,
the new method works on the FFT of size $n/m$ instead of $2n/m$ with KS method.



In this paper,
We use the term ``Frobenius partition'' for the set $\Sigma_{\omega}$ which partitions
the larger set $\Omega_n$ under the Frobenius map.

\subsection{Our Contributions}

A consequent problem arises upon \cite{VLL2017:carrylesspolymul}: 
Under a particular FFT, how to design a Frobenius partition 
resulting an efficient multiplier for bit-polynomials ?
For additive FFT, Li \emph{et al.}\cite{LCKCY2018:FAFFT} showed
two applicable Frobenius partitions for different applications.
One of the partition is for multiplying bit-polynomials of large degree
in modern computer.
However, they did not mention a concrete process for evaluating polynomials
at the particular partition. 

In this paper, we reformulate the Frobenius partition for additive FFT
over a simpler field and apply the partition to multiply bit-polynomials.
We first present the proof of correctness by counting the number of
deducible values and showing the enough number for evaluation and interpolation.
More importantly, 
we show how to fit the proposed partition into additive FFT
and the implementation techniques for a practical fast polynomial multiplier.




\section{Preliminaries}
\label{sec:preliminary}

\subsection{Multiplication of bit-Polynomials}
\label{sec:mul-poly-fft}
In this section,
we discuss the method for multiplying bit-polynomials of large degree.
It is well known that 
the multiplication can be done with FFT for evaluating polynomials
\cite{Cormen:2009:IntroAlgo}\cite{Gathen:2013:MCA:2512973}.


Given two polynomials $A(x) = a_0 + \cdots + a_d x^d$
and $B(x) = b_0 + \cdots + b_d x^d \in \GF{2}[x]_{\leq d}$,
represented in bit sequence of length $d+1 = \frac{n}{2}$ and
$n$ is a power of $2$\footnote{or we pad zero for the missing coefficients.},
we can calculate the product $C(x) = A(x) \cdot B(x)$ by evaluation and interpolation
as follows:
\begin{enumerate}
\setcounter{enumi}{-1}
\item Change the coefficient ring of the polynomials $A, B \in \GF{2}[x]_{< \frac{n}{2}}$
 from $\GF{2}$ to $\GF{2^m}$ where $2^m \geq n$. 
 It is actually nothing to do in this step. 
 However, we have conceptually $A, B \in \GF{2^m}[x]_{< \frac{n}{2}}$ with all 1-bit coefficients
 in $\GF{2^m}$, and there are enough points in $\GF{2^m}$ for evaluations.
\item Evaluate the $A$ and $B$ at $n$ points in $\GF{2^m}$ with FFTs.
\item Perform pointwise multiplications for the $n$ evaluated values.
\item Interpolate the values back to $C \in \GF{2^m}[x]_{< n}$,
 and then change the ring of coefficients such that $C \in \GF{2}[x]_{<n}$.
\end{enumerate}
The complexity of the polynomial multiplication is the same as the FFT in use.

\paragraph{Frobenius partitions of evaluated points}
In 2017, Van der Hoeven and Larrieu\cite{DBLP:conf/issac/HoevenL17} showed,
while evaluating polynomials over extending fields,
Forebenius map can derive more values of polynomials from less evaluated points.
Let $C \in \GF{2}[x]_{< n}$ and $\phi_2$ be the Frobenius map(square) over $\GF{2}$.
We can also apply $\phi_2$ to elements in extending fields, i.e.,
$\phi_2 : a \in \GF{2^m} \mapsto a^2$.
We note
\begin{equation}
\label{eq:Frobenius}
C( \phi_2(a) ) = \phi_2( C(a) ) \enspace,
\end{equation}
which means the value of $C$ at point $\phi_2(a)$ can be derived from
the value $C(a)$ by computing $\phi_2( C(a) )$.

We can then evaluate $C$ at a set $\Sigma$ and derive other values of $C$ 
from the values at $\Sigma$.
Let $\phi_2( \Sigma )$ be the set generated by 
applying $\phi_2$ to all elements in $\Sigma$ and
$\phi_2^{\circ j}$ be the function applying $\phi_2$ for $j$ times.

\begin{definition}
While applying $\phi_2$ to $\Sigma$ continuously,
let the order of the operation $\phi_2$ for a set $\Sigma$ be the minimal number 
obtaining the identical $\Sigma$, i.e.,
$\text{Ord}_{\phi_2}(\Sigma) = j $ for $j$ is the minimal number in $\mathbb{N}$
such that $\Sigma = \phi_2^{\circ j}(\Sigma)$.
\end{definition}

\begin{definition}(Frobenius partition)
Given $\text{Ord}_{\phi_2}(\Sigma) = j $,
we call $\Sigma$ a partition of $\Omega$ under Frobenius map if
\[
\Omega = \Sigma \cup \phi_2(\Sigma) \cup \cdots  \cup \phi_2^{ \circ (j-1)}(\Sigma)
\]
and all $\Sigma,\phi_2(\Sigma),\ldots,$ and $\phi_2^{ \circ (j-1) }(\Sigma)$
are disjoint sets. 
\end{definition}
By (\ref{eq:Frobenius}), all the values of $C$ at $\Omega$ 
can be derived from the values at $\Sigma$.

\subsection{Cantor Basis Representation of Binary Fields}
\label{sec:gf-arithmetic}
In this paper, we use row vectors over $\GF{2}$ to represent the elements of
binary fields(extension fields of \GF{2}).
The vectors in the space $\GF{2}^m$ are represented as $m$-bits binary strings
or alternatively the binary form of numbers $< 2^m$.
We use a line over a symbol to represent its vector form.
Under this convention, we define $ \overline{v_i} := 2^i$ and
the vectors $ v_0 , v_1 , \ldots , v_{m-1} $ form a basis for $\GF{2}^m$.

%


\subsubsection{Cantor Basis for Finite Field as Linear Space}
\label{sec:gf-ver3}
While representing an elements in the binary field $\GF{2^m}$
as a vector in the linear space $\GF{2}^m$,
Cantor\cite{Cantor:1989} showed a basis constructing the field of $\GF{2^m}$
for $m$ is a power of $2$, i.e., $ m = 2^{l_m}$ for $l_m \in \mathbb{N}$.
Gao and Mateer later used a simple construction of the 
Cantor basis in \cite{gm-afft}.
In the construction,
the Cantor basis $(v_i)$ satisfies 
\begin{equation}
\label{eq:cantor:basis}
v_0=1, \quad
v_i^2 + v_i = v_{i-1} \text{ for } i>0 \enspace . 
\end{equation}

\begin{definition} With respect to the basis $(v_i)$, let $\alpha := \sum_{j=0}^{m-1} 
b_j \cdot v_j$ be the \emph{field element represented by $\alpha$ under Cantor basis},
and the binary expansion of $ \overline{\alpha} = \sum_{j=0}^{m-1} b_j \cdot 2^j$ with $b_j \in \{ 0,1\}$.
\end{definition}

\begin{definition}
\label{def:space}
With respect to the basis $(v_i)$,  its 
\underline{sequence of subspaces} are
\[
V_0 := \{0\}, \quad 
V_i:=\mathrm{span}\{ v_0, v_1, \ldots, v_{i-1}\} \text{ for } i>0  \enspace.
\]
\end{definition}
Recall that $V_k$ is a field with basis $(v_j)_{j=0}^{k-1}$
for $k = 2^m$ is power of $2$.

We note that $V_0 \subset V_1 \subset V_2 \subset \cdots$ and the basis for smaller
spaces is the same as a part of the basis for larger spaces.
Hence, w.r.t. Cantor basis,
we can arbitrarily transform the elements in a smaller space to a larger
space by padding zero to extra dimensions \emph{without any cost}.
We thus treat elements in smaller fields as elements in larger fields arbitrarily.

\subsubsection{Subspace Polynomials over Cantor Basis}
\label{sec:si}

We introduce subspace polynomials in this section.
In contrast to the monomial basis $(1,x,x^2,\ldots)$,
we can also form a basis for polynomials with subspace polynomials in the next section.

\begin{definition}
\label{def:si}
With respect to the basis $(v_i)$, its
\underline{subspace polynomials} $(s_i)$ are,
\[
 s_i(x) := \prod_{a \in V_i} (x-a) \enspace.
\]
\end{definition}
We have $\deg(s_i(x)) = |V_i| = 2^i$ since $\dim(V_i)=i$.

Cantor \emph{et al.}\cite{Cantor:1989}\cite{gm-afft} showed the following useful
properties for $s_i$:
\begin{itemize}
\item (linearity) $s_i(x)$ is linear, i.e., $s_i(x+y) = s_i(x) + s_i(y)$.
\item (two terms for fields) $s_i(x) = x^{2^i} + x$ iff $i$ is a power of $2$, i.e., $V_i$ is a field. 
\item (recursivity) $s_i(x) = s_{i-1}^2(x) + s_{i-1}(x) = s_1(s_{i-1}(x))$; $s_{i+j}(x)=s_i(s_j(x))$.
\end{itemize}

With $s_0(x)=x$ and $s_{i+1} = s_i^2 + s_i$, by induction,
we know $s_i$ contains only terms with coefficients $1$ and monomials $x^{2^j}$.
Moreover, if
$i=2^{k_0}+2^{k_1}+\cdots+2^{k_j}$, where $2^{k_0}<2^{k_1}<\cdots<2^{k_j}$, then we
can write
\begin{equation}
\label{eq:split:si}
s_i(x) = s_{2^{k_0}}(s_{2^{k_1}}(\cdots(s_{2^{k_j}}(x))\cdots)) \enspace.
\end{equation}
Therefore, every $s_i$ is a composition of functions which only has two terms.

Evaluating $s_i$ w.r.t. Cantor basis is fast.
Directly from Def.~\ref{def:si}, we know $\forall a \in V_i, s_i(a) = 0 $.
For computing $s_i(v_j)$ for $j \geq i$, we have
\begin{equation}
\label{eq:eval:si}
s_i(v_j) = s_{i-1}( s_1(v_j) ) = s_{i-1}( v_j^2 + v_j ) = s_{i-1}( v_{j-1}) = \cdots = v_{j-i} \enspace .
\end{equation}
Hence, the value
  of $s_i(\alpha)$ is $\overline{\alpha}$ shifted right by
  $i$ bits, or $s_i( \alpha ) = ( \overline{\alpha} \gg i )= \sum_{j=i}^{m-1} b_j 2^{j-i}$. 



\subsection{The Additive FFT}
\label{sec:addfft}
In this section,
we show how to efficiently evaluate a polynomial $f \in \GF{}[x]_{<n}$ 
at $n$ points in Cantor basis.
Again, assume $n = 2^{l_n}$ is a power of $2$.
The additive FFT of the form by Lin, Chung, and Han (or \addfft)\cite{lch-afft} 
requires that $f$ is represented in a particular basis, called \novelpoly basis.


 

\begin{definition}\label{def:novelpoly}
Given the Cantor basis $(v_i)$ for the base field and its 
subspace polynomials $(s_i)$,
define the \underline{\novelpoly basis} w.r.t.\ $(v_i)$ to be the polynomials $(X_k)$
\[
X_k(x):= \prod \left(s_i(x)\right)^{b_i} \quad 
\mbox{ where } k=\sum b_i \cdot 2^i \mbox{ with } b_i \in \{ 0,1\}\enspace .
\]
I.e.,
$X_k(x)$ is the product of all $s_i(x)$ where the $i$-th bit
of $k$ is set.
\end{definition}
Clearly, $X_{2^i}(x) = s_i(x)$.
Since $\deg(s_i(x)) = 2^i$, we have $\deg(X_k(x)) =k$.

%

\subsubsection{Basis Conversion for Polynomials}
\label{sec:basis-conversion}
We have to write the polynomial $f \in \GF{}[x]$ in the form
$f(x) = g(X) = g_0 + g_1 X_1(x) + \ldots + g_{n-1} X_{n-1}(x)$
to perform $\addfft$.

In \cite{auth256}, Bernstein and Chou convert $f(x)$ to $g(X)$ by finding the
largest $i$ such that $\deg( s_i )= 2^i < \deg (f )$, and then divide $f$ by
$s_i$ to form $f(x) = f_0(x) + s_i(x) f_1(x)$.  Recursively divide
$f_0$ and $f_1$ by lower $s_{i-1}$ and eventually express $f$
as a sum of non-repetitive products of the $(s_i)$, which is the
desired form for $g(X)$.  
Since the coefficients of $s_i(x)$ are always $1$ in Cantor basis,
the division comprises only $\xor$ operations.
Therefore the complexity of division by one $s_i$ depends on the number
of terms of $s_i$ 
and the complexity of the conversion is $O(n \log^n (\log^n)^2)$ field additions.


With Eq.(\ref{eq:split:si}),
Lin \emph{et al.} \cite{Lin:BasisCvt:2016} presented a basis conversion
in Cantor basis by dividing $f$ by $s_{i}$ where $i=2^k$ is power of $2$
and $s_i$ contains only 2 terms.
The resulted complexity for the conversion is $O(n \log^n \log(\log^n))$ $\xor$
operations for $f \in \GF{2}[x]_{< n}$.
We detail the conversion in Alg.~\ref{alg:changeBasis2}.
%
\begin{algorithm}[!h]
\SetKwFunction{vs}{VarSubs}
\SetKwFunction{cvt}{BasisCvt}
\SetKwInOut{Input}{input}
\SetKwInOut{Output}{output}
\cvt{$f(x) \in \RR{}[x] $} : \\
\Input{ $ f(x) = f_0 + f_1 x + ... + f_{n-1} x^{n-1} \in \RR{}[x]$ .}
\Output{ $ g(X) = g_0 + g_1 X_{1}(x) + ... + g_{n-1} X_{n-1}(x) \in \RR{}[x]$ .}
\BlankLine
\lIf { $ \deg( f(x) ) \leq 1 $ }
{
  return $g(X) \gets f_0 + X_1 f_1$
}
Let $ i \gets \text{Max}(2^k)$ where $k \in \mathbb{N} \quad \text{s.t. } \deg (s_{i}(x)) \leq \deg (f(x)) $ .\\
Compute $h(x,y) = \hat{h}(y) = h_0(x) + h_1(x) y + \cdots + h_{m-1}(x) y^{m-1} \in \RR{}[x][y] $
  s.t. $ f(x) = h(x, s_i(x)) $.
   Here $h_0(x),\ldots,h_{m-1}(x) \in \RR{}[x]_{< 2^i}$ . \\
Compute $h'(Y) = q_0(x) + \cdots + q_{m-1}(x) X_{(m-1)\cdot 2^{k}} \gets $ \cvt{ $\hat{h}(y)$ } . \\
Compute $g_i(X) \gets $ \cvt{ $q_i(x)$ }  for all coefficients $q_i(x)$ of $h'(Y)$. \\
\Return $ g(X) = g_0(X) + g_1(X) X_{2^k} + ... + g_{n-1}(X) X_{(m-1)\cdot 2^k} $ \\
\caption{Basis conversion: monomial to \novelpoly w.r.t\ Cantor.}
\label{alg:changeBasis2}
\end{algorithm}

\subsubsection{The Butterflies}
For a polynomial $f = g(X) \in \GF{2^m}[x]_{<n}$ in the $\novelpoly$ basis, 
we can efficiently evaluate $f$ at the set $\alpha + V_{l_n}$, 
where $\alpha \in \GF{2^m}$, $n=2^{l_n}$, and $\alpha+V_{l_n} := \{ \alpha + u : u \in V_{l_n} \}$,
 with a ``Butterfly'' process,
denoted as $\lchbtfy$. \footnote{ Note that, in \cite{lch-afft}, the authors call the $\lchbtfy$ an FFT.}

\begin{algorithm}[!tbh]
\SetKwFunction{nfft}{\lchbtfy}
\SetKwInOut{Input}{input}
\SetKwInOut{Output}{output}
\nfft{$g(X) = f(x) \in \GF{2^m}[x]_{< n}, \alpha \in \GF{2^m} $} : \\
\Input{ $ g(X) = g_0 + g_1 X_1(x) + ... + g_{n-1} X_{n-1}(x) \in \GF{2^m}[x]_{< n}$ .\\
        an extra scalar: $ \alpha \in \GF{2^m} $ .\\}
\Output{ a list: $ [ f( 0 + \alpha ),f( 1 + \alpha ), \ldots , f( (\overline{n-1}) + \alpha)  ] $ , which
   is the evaluation of $f$ at $\alpha + V_{l_n}$ .}
\BlankLine
\lIf { $ \deg( g(X) ) = 0 $ }{ return $[ g_0 ]$ }
Let $i \gets \lceil \log_2^n \rceil - 1 $ s.t. $ \deg( s_i(x)) = \deg(X_{2^i}(x)) = 2^i < n $ . \\
Let $ g(X) = p_0(X) + X_{2^i} \cdot p_1(X) = p_0(X) + s_i(x) \cdot p_1(X) $. \\
Compute $h_0(X) \gets p_0(X) + s_i(\alpha) \cdot p_1(X) $. \\
Compute $h_1(X) \gets h_0(X) + s_i(v_i) \cdot p_1(X) $. \\ 
\Return $[$ \nfft{$  h_0(X) , \alpha $},\nfft{$ h_1(X) , v_i + \alpha $} $]$
\caption{$\lchbtfy$ w.r.t. $\novelpoly$ basis.}
\label{alg:fft}
\end{algorithm}

We detail the $\lchbtfy$ in the Algo.~\ref{alg:fft}.
It is a typical divide-and-conquer process that
the polynomial $f = g(X) $ can be expressed 
as two half-sized polynomials $p_0(X)$ and $p_1(X)$ with
$g(X) = p_0(X) + X_{2^{l_n - 1}} (x) \cdot p_1(X)$.
For evaluating $f$ at a set $V_{i+1}$ where $i+1 = l_n$,
we also divide $V_{i+1}$ into the two half-sized sets $V_i$ and 
$V_{i+1}\backslash V_{i} = v_i + V_{i}$. 
Since $s_i(x) = X_{2^i}(x)$ is linear, 
evaluations at $V_{i}+v_{i}$ share common computations
with the evaluations at $V_{i}$.
%
%

In the Algo.~\ref{alg:fft}, line~5 and 6 perform the actual computation -- the so-called butterfly.
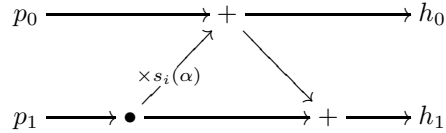
\[
\xymatrix{
   **[l] p_0 \ar[rr] & &  + \ar[rr] \ar[rd] &  & h_0 \\
   **[l] p_1 \ar[r]  & \bullet \ar[ru]|(.4){ \times s_i(\alpha) } \ar[rr] & & + \ar[r] & h_1 \\
}
\]
One butterfly comprises two field additions and only one field multiplication
since $s_i(v_i)=1$ in Cantor basis.

Although line~7 indicates the recursion, we actually program the recursions 
into many layers of butterflies.
There are $n/2$ butterflies, corresponding to the length of divided polynomials, in each layer
and $l_n$ layers, corresponding to the depth of recursion, in total.
Through the iterative style of program, we can optimize the $\lchbtfy$ among several layers(recursions), 
and it is also applied to the $\basiscvt$.

We remark at last that the Algo.~\ref{alg:fft} expects that length of polynomials
and size of evaluated points are the same.
We will evaluate polynomials of larger degrees at a smaller size of set
in Sec.~\ref{sec:truncated:fft}.




\subsubsection{The $\addfft$ Algorithm}
The complete $\addfft$ is show in Algo.~\ref{alg:addfft}.
The algorithm performs $\basiscvt$ to convert the basis of polynomials and
then use the $\lchbtfy$ to evaluate polynomials in the $\novelpoly$ basis.
Inverse additive FFT, or $\iaddfft$, simply performs the $\lchbtfy$ and $\basiscvt$ reversely.

\begin{algorithm}[!tbh]
\SetKwFunction{afft}{\addfft}
\SetKwInOut{Input}{input}
\SetKwInOut{Output}{output}
\afft{ $ f(x) \in \GF{2^m}[x]_{< n}, \alpha \in \GF{2^m} $} : \\
\Input{ $ f(x) = f_0 + f_1 x + ... + f_{n-1} x^{n-1} \in \GF{2^m}[x]_{< n}$ .\\
        an extra scalar: $ \alpha \in \GF{2^m} $ .\\ }
\Output{ The values for evaluating $f$ at $\alpha + V_{l_n}$ .}
\BlankLine
Compute $g(X) \gets \basiscvt( f(x) ) .$ \\
\Return $ \lchbtfy( g(X) , \alpha ) . $ \\
\caption{The $\addfft$ algorithm}
\label{alg:addfft}
\end{algorithm}


\section{The multiplication with Froebenius partitions and additive FFT}
\label{sec:frobenius:bijection}

In this section, we apply the technique of Frobenius partitions to
the $\addfft$ and show an efficient multiplier for 
bit-polynomials with the modified $\addfft$.

\subsection{The partition of evaluated points}

Given a polynomial $A \in \GF{2}[x]$ for $\deg(A) = n-1$ and $n=2^{l_n}$, we aim
to design a set $ \Sigma \subset \GF{2^m}$ for $m=2^{l_m}$ in Cantor basis
such that we can derive $n$ values of $A(x)$ from the values at $\Sigma$.

Before we define the $\Sigma$, we first discuss the order of $\phi_2$ for the 
basis elements.
Given $\phi_2$ is the square operation over $\GF{2^m}$, we have
$\phi_2(v_0) = v_0$, and $\phi_2(v_i) = v_i^2 = 
v_i + v_{i-1}$ for $i > 0$ in Cantor basis.
In \cite{LCKCY2018:FAFFT}, Li \emph{et al.} show the order of $\phi_2$ 
\begin{equation}
\label{eq:order}
\text{Ord}_{\phi_2}(v_i) =  2 \cdot 2^{ \lfloor \log_2 i \rfloor} \quad \text{ for } i > 0
\end{equation}
w.r.t. Cantor basis.
Hence, $\text{Ord}_{\phi_2}( v_i ) = m $ is the maximum order 
for $ v_i \in \{ v_{m/2}, \ldots , v_{m-1} \} $.

We design a set with maximum order of $\phi_2$:
\begin{equation}
\label{eq:the:partition}
\Sigma := v_{l+m/2} + V_l  \quad \text{ where } l = (l_n - l_m) < m/2 \text{ and } l \ge 0 \enspace ,
\end{equation}
and size of the set is $ n_p := |\Sigma| = |V_l| = 2^l = n/m $.
For the order, we know first that $\phi_2$ maps $V_l$ to the same $V_l$, i.e., $\phi_2(V_l) = V_l$.
 This can be seen by induction. 
 Clearly, $\phi_2(V_0) = V_0$. Given $\phi_2(V_i) = V_i$, we have
 \[
  \phi_2(V_{i+1}) = \phi_2(v_i + V_i) \cup \phi_2(V_i)
  = ((v_i + v_{i-1})+ V_i ) \cup V_i = (v_i + V_i) \cup V_i = V_{i+1} \enspace .
 \]
Hence, $v_{l+m/2}$ decides the order for $\Sigma$ and 
naively, $\text{Ord}_{\phi_2}(v_{l+m/2}) = m$ from Eq.~\ref{eq:order}.
However, since $\Sigma = v_{l+m/2} + V_l$, we have to deal with the effect from $V_l$.
While applying $\phi_2$ to $v_{l+m/2}$ for $j$ times,
let the vector $ \phi_2^{\circ j}( v_{l+m/2} ) = a + b $ where $a \in V_l$ is equal to 
least $l$ dimensions of $ \phi_2^{\circ j}( v_{l+m/2} )$ and $b$ is the remainder.
Then $V_l + a = V_l $ since $a \in V_l$.
And, by omitting the least $l$ dimensions of $b$,
the order for the higher parts of $\phi_2^{\circ j}( v_{l+m/2} )$ is still $m$
\[
\text{Ord}_{\phi_2}(\bar{b}\gg l ) = \text{Ord}_{\phi_2}(v_{m/2}) = m \enspace .
\]
Hence, $\text{Ord}_{\phi_2}(\Sigma) = m$.


By continuously applying $\phi_2$ to $\Sigma$, we define a superset of $\Sigma$
\begin{equation}
\label{eq:superset}
\Omega := \Sigma \cup \phi_2(\Sigma) \cup \phi_2^{\circ 2}  ( \Sigma ) \cup \cdots \cup 
\phi_2^{\circ (m-1)}(\Sigma) \enspace .
\end{equation}
\begin{proposition}
\label{lemma:the:partition}
$\Sigma$ is a Frobenius partition of $\Omega$ and $|\Omega| = n $.
\end{proposition}
While continuously applying $\phi_2$ to $\Sigma$, from the discussion of the order for $\Sigma$, 
the $V_l$ absorbs $a$ parts of $\phi_2^{\circ j}( v_{l+m/2} )$,
and only the $b$ parts of $\phi_2^{\circ j}( v_{l+m/2} )$ changes.
Hence, $\phi_2^{\circ j}( \Sigma)$ are disjoint sets for $j<m$, 
and $\Sigma$ is a Frobenius partition of $\Omega$.
And the size
$|\Omega| = \text{Ord}_{\phi_2}( \Sigma ) \cdot |\Sigma| = m \cdot \frac{n}{m} = n$.


Now we define a linear map $ E_{\Sigma} : \GF{2}[x]_{< n} \rightarrow \GF{2^{m}}^{n_p}$,
evaluating a polynomial $A \in \GF{2}[x]_{< n}$ at the partition $\Sigma$, i.e.,
\begin{equation}
\label{eq:eval:at:sigma}
E_{\Sigma}: A(x) \mapsto \{ A( v_{l+m/2} + u) : u \in V_l \} \enspace .
\end{equation}
%
Clearly, $E_{\Sigma}$ evaluates $A(x)$ at $n_p = n/m $ points
which are fewer than number of coefficients $n$.
However, since the points are in $\GF{2^{m}}$, 
the size of input and output space are the same $n = m\cdot n/m$ bits.
\begin{proposition}
\label{lemma:the:bijection}
$E_{\Sigma}$ is a bijection between $\GF{2}[x]_{<n}$ and $\GF{2^{m}}^{n_p}$.
\end{proposition}
$E_\Omega$ is clearly a bijection between $\GF{2}[x]_{<n}$ and $\GF{2^{m}}^{n}$ since
its the evaluation at $n$ points.
$\Sigma$ can derive full $\Omega$ with the linear operator $\phi_2$ and vice versa.
By Eq.(\ref{eq:Frobenius}), $E_\Omega$ can be derived from $E_\Sigma$ with linear operator $\phi_2$.

Given the field $\GF{2^m}$, the size of $\Omega$ bounds the possible length
of polynomials $n$.
From Prop.~\ref{lemma:the:partition}, 
$|\Omega|= \text{Ord}_{\phi_2}( \Sigma ) \cdot |\Sigma| = m \cdot |\Sigma|$.
Since the maximum $|\Sigma| = |V_{\frac{m}{2} - 1}| = 2^{\frac{m}{2} - 1}$
for $l<m/2$ in Eq.~\ref{eq:the:partition}, 
the maximum $|\Omega| = m \cdot 2^{ \frac{m}{2}-1} = \frac{m}{2} \cdot 2^{m/2}$.
Therefore, given $\GF{2^m}$, $ n < \frac{m}{2} \cdot 2^{m/2} $ is the 
maximum supported length of polynomials.


\subsection{Truncated Additive FFT}
\label{sec:truncated:fft}

To perform $E_{\Sigma}$, we have to evaluate a polynomial $A$ of length $n$ at a $n/m$ points
with the $\addfft$.
Since the $\basiscvt$ only depends on the polynomial, we only have to adjust the $\lchbtfy$.
Recalling that Algo.~\ref{alg:fft} outputs the values of $A$ at points 
$\{ \alpha , \alpha + 1 , \ldots , \alpha + (n-1) \}$,
we can simply truncate the computation of unnecessary outputs for more efficiency
--- the so-called truncated FFT.

\begin{figure}[h]
\caption{The $\lchbtfy$ of 2 layers truncates half contents after first layer. }
\label{fig:trunc:fft}
\[
\xymatrix{
   **[l] g_0 \ar[rr] \ar@{--}[]+<1.5em,1em>;[ddd]+<1.5em,-1em> &  & + \ar[rrr] \ar[rdd] & \ar@{--}[]+<1.5em,1em>;[ddd]+<1.5em,-1em> &  & + \ar[rr] \ar[rd] & \ar@{--}[]+<1.5em,1em>;[ddd]+<1.5em,-1em> & f(0+\alpha) \\
   **[l] g_1 \ar[rr] &  & + \ar[rr] \ar[rdd] & & \bullet \ar[ru]|(.4){ \times s_0(\alpha) } \ar[rr] & & + \ar[r] & f(1+\alpha) \\
   **[l] g_2 \ar[r] & \bullet \ar[ruu]|(.3){\times s_1(\alpha)} \ar[rr] & & +  \ar[r]  & \bullet  & & & \\
   **[l] g_3 \ar[r] & \bullet \ar[ruu]|(.3){\times s_1(\alpha)}  \ar[rr] & & + \ar[r]  & \bullet  & & & \\
}
\]
\end{figure}
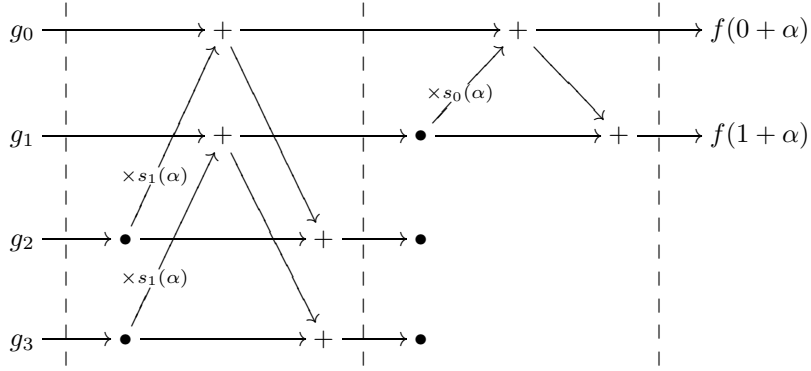
Fig.~\ref{fig:trunc:fft} shows an example of truncated FFT for evaluating
a degree-3 polynomial $f(x) = g_0 + g_1 X_1 + g_2 X_2 + g_3 X_3$ at two points $\{ \alpha , \alpha + 1 \}$. 
There are 2 layers(recursions) of butterflies in the computation of $\lchbtfy$.
We can truncate the half contents after the first layer since only $2$ values are required.

To perform $E_{\Sigma}$, we therefore pretend to evaluate $A$ at a larger set $v_{l+m/2} + V_{l_n}$ with
the $\lchbtfy$. 
Although $v_{l+m/2} + V_{l_n} \neq \Omega$, the size of two sets are the same.
However, after $l_m$ layers of butterflies, the computations for the values at $\Sigma$
aggregates to the first $n/m$ parts of the layer, and we can thus truncate the rest.


\subsection{Encoding: the First $ l_m $ Layers of the Truncated FFT}

While performing the first $l_m$ layers of the $\lchbtfy$ in $E_{\Sigma}$,
the temporary results expand by $m$ times since the inputs are 1-bit data and
the multipliers in butterflies are $m$-bits elements of $\GF{2^m}$.
However, since we will truncate the temporary results to the factor of $1/m$
after $l_m$ layers of butterflies,
the space requirement balances after the data truncation.
Hence, we design a process, the $\enc$\footnote{and its reverse process, the $\dec$.},
to prevent the expansion from the first $l_m$ layers of butterflies.

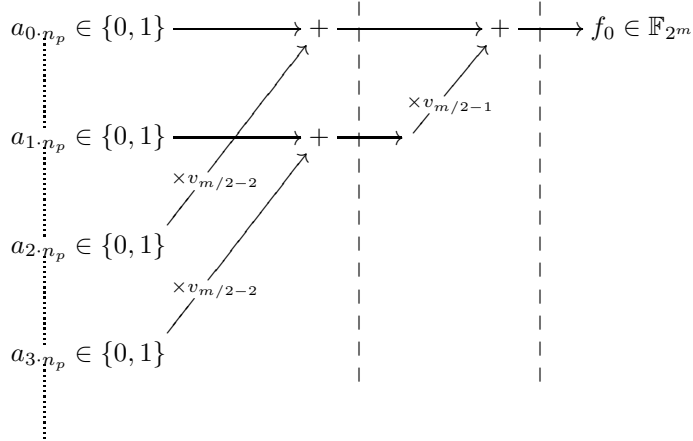
\begin{figure}[h]
\caption{The first temporary result $f_0$ after 2 layers of butterflies.
The multipliers in the butterflies are 
$s_{l+2}(v_{l+m/2}) = v_{m/2-2}$ and $s_{l+1}(v_{l+m/2}) = v_{m/2-1}$
for the first and second layers respectively.}
\label{fig:trunc:2l}
\[
\xymatrix{
   **[l] a_{0 \cdot n_p} \in \{0,1\} \ar@{.}[]-<4em,0.4em>;[d]-<4em,-0.4em> \ar[r] & + \ar[rr] \ar@{--}[]+<1.5em,1em>;[ddd]+<1.5em,-1em> &    & + \ar[r] \ar@{--}[]+<1.5em,1em>;[ddd]+<1.5em,-1em> & f_0 \in \GF{2^m} \\
   **[l] a_{1 \cdot n_p} \in \{0,1\} \ar@{.}[]-<4em,0.4em>;[d]-<4em,-0.4em> \ar[r]   & + \ar[r] & \ar[ru]|(.3){ \quad \times v_{m/2-1}} & & \\
   **[l] a_{2 \cdot n_p} \in \{0,1\} \ar@{.}[]-<4em,0.4em>;[d]-<4em,-0.4em>  \ar[ruu]|(.3){ \quad \times v_{m/2 - 2}} & &               & & \\
   **[l] a_{3 \cdot n_p} \in \{0,1\} \ar@{.}[]-<4em,0.4em>;[d]-<4em,-0.4em>  \ar[ruu]|(.3){ \quad \times v_{m/2 - 2}} & &              & & \\
   & & & & & \\
}
\]
\end{figure}

We show the computations of the first temporary result
\[
f_0 = a_0 + s_{l+1}(v_{l+m/2}) \cdot a_{n_p} + s_{l+2}(v_{l+m/2}) \cdot a_{2\cdot n_p}
+ s_{l+1}(v_{l+m/2}) \cdot s_{l+2}(v_{l+m/2}) \cdot a_{3 \cdot n_p} 
\]
for 2 layers of the $\lchbtfy$ in Fig.~\ref{fig:trunc:2l}.
The $\lchbtfy$ evaluates a polynomial $A(X) = a_0 + a_1 X_1 + \cdots + a_{n-1} X_{n-1} \in \GF{2}[x]_{<n}$
at the set $v_{l+m/2}+V_{l_n}$ where $2^l = n_p = n/m$.
If $m=4$, for example, the storage space for $f_0 \in \GF{2^4}$ equals its 4 contributors
$(a_0, a_{n_p}, a_{2\cdot n_p},a_{3\cdot n_p})$.

For the first $l_m$ layers, the multipliers in the butterflies are
the evaluation of $( s_{l+ l_m} , \ldots , s_{l+1} )$ at the same point $v_{l+m/2}$.
With Eq.~(\ref{eq:eval:si}), we can calculate the multipliers in reverse order of layers
\[
( s_{l+ 1}( v_{l+m/2} ) , \ldots , s_{l+ l_m}( v_{l+m/2} ) ) = ( v_{m/2 - 1} , \ldots , v_{m/2 - lm} ) \enspace .
\]
We note that the multipliers are independent of $l$, i.e.,
the multipliers are always $( v_{m/2 - 1} , \ldots , v_{m/2 - lm} )$ for a given $m$.

We can further analyze the multipliers for distinct inputs.
The multiplier for $j$-th input is
\begin{equation}
\label{eq:encode:multipliers}
r_j = \prod_{k=0}^{l_m-1} (v_{m/2-1-k}) ^{j_k} 
       \quad,\text{for } j = j_0 + j_1 \cdot 2 + \cdots + j_{l_m-1} \cdot 2^{l_m - 1} \enspace .
\end{equation}
Hence, we can show $ \enc : A(X) \mapsto  f(X) = f_0 + \cdots + f_{n_p-1}X_{n_p-1} \in \GF{2^m}[x]_{<n_p}$ 
in the form of vector-matrix production for all results:
\[ 
f_i = \sum_{j=0}^{m-1} a_{j \cdot n_p + i} \cdot r_j =   
\begin{bmatrix}
    a_{0 \cdot n_p}  &
    a_{1 \cdot n_p}  &
    \ldots  &
    a_{(m-1) \cdot n_p} \\
  \end{bmatrix}
  \cdot
  \begin{bmatrix}
     r_0 \in \GF{2}^m \\
     r_1 \in \GF{2}^m \\
     \vdots \\
     r_{m-1} \in \GF{2}^m \\
  \end{bmatrix} \enspace .
\]



The algorithm of the $\enc$ is listed in Algo.~\ref{alg:encoding}.
\begin{algorithm}[!tbh]
\SetKwFunction{encode}{\enc}
\SetKwInOut{Input}{input}
\SetKwInOut{Output}{output}
\encode{ $ A(X) \in \GF{2}[x]_{< n} $} : \\
\Input{ $ A(X) = a_0 + a_1 X_1 + \cdots + a_{n-1} X_{n-1} \in \GF{2}[x]_{< n}$ .}
\Output{ $ f(X) = f_0 + \cdots + f_{n_p-1} X_{n_p -1 } \in \GF{2^m}[x]_{< n_p} $ where $n_p = n/m $. }
\BlankLine
\For  { $ i \in \{ 0,\ldots,n_p-1\} $ } {
Collect  $( a_{ i+ 0\cdot n_p}, a_{ i+ 1\cdot n_p}, \ldots , a_{ i+ (m-1)\cdot n_p} ) $. \\
Compute  $f_i \gets \sum_{j=0}^{m-1} a_{j \cdot n_p + i} \cdot r_j $ .\\ 
}
\Return $( f_0, f_1 ,  \ldots , f_{n_p-1} ) $ .
\caption{The \enc }
\label{alg:encoding}
\end{algorithm}

For collecting the $m$ bits inputs $(a_{j \cdot n_p + i})_{j=0}^{m-1}$ efficiently
in computer, 
we actually fetch $m$ machine words of length $w$-bits instead of $m$ separated bits.
The component $a_{j \cdot n_p + i}$, for example, locates in the $i$-th bit of the $j$-th word.
With an $m \times w$ matrix transpose, 
we can collect the inputs(Line~3) for $w$ continuous indexes of the loop.
The Line~4 in Algo.~\ref{alg:encoding} can also be parallelized with a bit-matrix multiplication.

\subsection{The Algorithm of Multiplication}

We show the algorithm for multiplying bit-polynomials in algo.~\ref{alg:polymul}.
It is basically the general multiplication in Sec.~\ref{sec:mul-poly-fft}
with a modifier $\addfft$ comprising $\basiscvt$, $\enc$, and $\lchbtfy$.

\begin{algorithm}[!tbh]
\SetKwFunction{polymul}{ \ensuremath{\texttt{FP\_Polymul}\xspace} }
\SetKwInOut{Input}{input}
\SetKwInOut{Output}{output}
\polymul{ $ A(x), B(x) \in \GF{2}[x]_{< n/2} $} : \\
\Input{ $ A(x) = a_0 + \cdots + a_{n/2-1} x^{n/2-1} \in \GF{2}[x]_{< n/2}$ . \\ 
  $ B(x) = b_0 + \cdots + b_{n/2-1} x^{n/2-1} \in \GF{2}[x]_{< n/2}$ . }
\Output{ $ C(x) = A(x) \cdot B(x) = c_0 + \cdots + c_{n-1} x^{n-1}  \in \GF{2}[x]_{< n}$ }
\BlankLine
Compute $A(X) \in \GF{2}[x]_{<n/2} \gets \basiscvt( A(x) ) .$ \\
Compute $ f(X) \in \GF{2^m}[x]_{<n_p}  \gets \enc( A(X) )$. \\
Compute $ [\hat{a}_1,\ldots, \hat{a}_{n_p}] \in \GF{2^m}^{n_p} \gets \lchbtfy( f(X) , v_{l + m/2} ) $. \\
Compute $ B(X) \in \GF{2}[x]_{<n/2} \gets \basiscvt( B(x) ) .$ \\
Compute $ g(X) \in \GF{2^m}[x]_{<n_p}  \gets \enc( B(X) )$. \\
Compute $ [\hat{b}_1,\ldots, \hat{b}_{n_p}] \in \GF{2^m}^{n_p} \gets \lchbtfy( g(X) , v_{l + m/2} ) $. \\
Compute $ \hat{C} = [ \hat{c}_1 \gets \hat{a}_1 \cdot \hat{b}_1 , \ldots , \hat{c}_{n_p} \gets \hat{a}_{n_p} \cdot \hat{b}_{n_p} ] $. \\
Compute $ h(X) \in \GF{2^m}[x]_{<n_p} \gets \ilchbtfy( \hat{C} , v_{l + m/2} ) $. \\

Compute $ C(X) \in \GF{2}[x]_{<n} \gets \dec( h(X)) $. \\
Compute $ C(x) \in \GF{2}[x]_{<n} \gets \ibasiscvt( C(X) ) $. \\
\Return $C(x)$. \\
\caption{The multiplication of bit-polynomials}
\label{alg:polymul}
\end{algorithm}

Here we sum up the modified $\addfft$ process:
To evaluate a polynomial $A(x) \in \GF{2}[x]_{<n}$ at $n_p$ points $\Sigma$ in $\GF{2^{m}}$,
we first perform the $\basiscvt$ 
on $A(x)$ for $A(X)$ in $\novelpoly$. 
Then we treat each coefficient of $A(X)$ as an element in $\GF{2^{m}}$
and pretend to perform the $\lchbtfy$ 
at points $v_{l+m/2} + V_{l_n}$.
The $\enc$ process actually perform the virtual $\lchbtfy$ for the first $ l_m $ layers of butterflies
and truncate the temporary results to the first $1/m$ fraction.
We then start a real $\lchbtfy$ on the results of the $\enc$.
The $\lchbtfy$ evaluates a polynomial in $\GF{2^{m}}[x]_{<n_p}$ at $\Sigma$.


\section{Implementation}
\label{sec:imple}

In this section,
we discuss the implementations for previous algorithms in modern computers.
We choose the parameters $m = 64 $ and $128$
for the efficiency and supporting larger length of polynomials respectively.
The corresponding implementations, over $\GF{2^{64}}$ and $\GF{2^{128}}$,
apply to multiply bit-polynomials of $\GF{2}[x]_{<32 \cdot 2^{32}}$ and $\GF{2}[x]_{<64 \cdot 2^{64}}$ respectively.


\subsection{Memory Access Model}

We first discuss about our memory access model for the $\basiscvt$ and $\lchbtfy$.

\paragraph{Basis conversion:}
We focus on reducing the number of memory access for optimizing the $\basiscvt$.
For process of only simple $\xor$ operations,
Albrecht \emph{et al.}\cite{DBLP:journals/toms/AlbrechtBH10} 
reported that the number of memory access is the critical concern 
while multiplying matrices over $\GF{2}$.
The $\basiscvt$ face the same situation. 

For reducing the number of memory access,
we combine several layers of operations together in algo.~\ref{alg:changeBasis2}.
It is possible since
the alog.~\ref{alg:changeBasis2} always $\xor$
coefficients of higher degree to coefficients of lower degree.
While same coefficients of lower degree
gather coefficients from higher degree among several layers,
we can combine the accumulations among layers.
This optimization effectively reduces the number of memory write.


\paragraph{The Butterfly:}

The memory access model in $\lchbtfy$ focuses on hiding
the time for memory access behind the computations.
Instead of memory bound in $\basiscvt$,
the $\lchbtfy$ multiplies elements over finite fields and thus
is occupied with heavy computations.
For hiding the memory access, first, we change the order of butterflies performed
to keep as more data in CPU cache as possible.
We divide the butterflies in one layer into batches which fits for the size of cache
and perform butterflies throughout all layers in the same batch to keep a higher hit rate.
Second, we pre-fetch the data for next computation before acting multiplications.
The pre-fetch hints the CPU to move the data to fastest cache covertly
behind the ongoing task, and we can thus reduce the latency for next reading.


\subsection{The Vector Instruction Set}

Besides the optimization in memory access,
we also target on the hardware instructions
that increase the efficiency for the algebraic objects.

We implement our software in the typical SIMD(single-instruction-multiple-data) instruction set.
The most popular SIMD instruction set nowadays is AVX2(Advanced Vector Extensions)\cite{intel-isa},
providing 256-bit \texttt{ymm} registers on x86 platforms.
We especially rely on the table-lookup($\pshufb$) and 
carryless multiplication($\pclmul$) instructions.

\paragraph{SIMD Table-lookup Instruction} \pshufb takes two 16-byte sources which one is a lookup table of 16
bytes $\bm{x} = (x_0,\, x_1,\allowbreak \ldots,\allowbreak \, x_{15})$
and the other is 16 indices
$\bm{y} = (y_0,\, y_1,\allowbreak \ldots,\allowbreak \, y_{15})$.  The 16-byte
result of ``\pshufb $\bm{x},\bm{y}$'' at position $i$ is $x_{y_i\!\!\!\mod 16}$ if
$y_i\ge 0$ and $0$ if $y_i<0$.  
\vpshufb of AVX-2 simply performs two copies of \pshufb in one instruction.
\paragraph{Carryless Multiplication}
$\pclmul$ performs the carryless multiplication of 2 64-bits polynomials,
i.e., $\pclmul : \GF{2}[x]_{<64} \times \GF{2}[x]_{<64} \rightarrow \GF{2}[x]_{<127}$.
This is unfortunately not a SIMD instruction despite the high efficiency in multiplication.

\subsection{Finite Field Arithmetic}

In this section, we discuss the representations of fields in polynomial form as well as its
corresponding multiplications for $m=64$ and $m=128$.
Although we design the algorithm in vector representations of Canto basis,
we actually use the polynomial representation while multiplying elements in fields
for the dedicated HW instructions $\pclmul$.
Hence, we have to change the representations of fields from Cantor basis
to the polynomial form before acting multiplications, and
some tables are prepared for changing the representations.
However, since the inputs are in $\{ 0, 1 \}$ that can be presumed in polynomial form,
we actually perform the change representations only for the constants in $\lchbtfy$.
%

For the multiplication over $\GF{2^{64}}$, 
Lemire and Kaser \cite{LemireK16} presented an efficient multiplication
under the representation
\[
\GF{2^{64}} := \GF{2}[x]/\left( x^{64} + x^4 + x^3 + x + 1\right) \enspace .
\]
To multiply elements in $\GF{2^{64}}$, one $\pclmul$ multiplying 2 degree-63
bit-polynomials to a degree-126 polynomial and then one $\pclmul$ reduces the parts of
degree-64 to 126 back to a remainder of degree-66. One $\pshufb$ finishes the reduction
for the degree-64 to 66.

For $\GF{2^{128}}$, we choose the same representation as AES-GCM:
\[
\GF{2^{128}} := \GF{2}[x]/\left( x^{128} + x^7 + x^2 + x + 1\right) \enspace .
\]
In our implementation, the multiplication over \GF{2^{128}} costs 5 \pclmul(3 for multiplying $128$-bit
polynomials with Karatsuba's method and 2 for reducing the $256$-bit result
back to $128$ bits with linear folding).
More details about multiplications over \GF{2^{128}} can be found in \cite{pclmul:gcm}.

\subsection{Matrix Transpose with Vector Instruction Set}
\label{sec:matrix:transpose}

We perform the  matrix transpose with the techniques from \cite{hackersdelight}.
For implementing it in the SIMD manner,
Van der Hoeven \emph{et al.}\cite{VLL2017:carrylesspolymul} and 
Chen \emph{et al.}\cite{DBLP:journals/corr/abs-1708-09746}
had showed similar techniques for bit- and byte-matrix in AVX-2 instruction set.
We depict the methods in this section for the completeness.

The method for matrix transpose in \cite{hackersdelight} is a divide-and-conquer method.
For transposing $ M = \begin{bmatrix} A & B \\ C & D \\ \end{bmatrix} $, we first rearrange
the contents of $M$ to $ \begin{bmatrix} A & C \\ B & D \\ \end{bmatrix} $ and then perform
the same process to all 4 sub-matrices.

The interpretation of data plays an important role for transpose in a SIMD instruction set.
Given $A = \begin{bmatrix} a_0 & a_1 \\ a_2 & a_3 \\ \end{bmatrix}$ is an $2\times 2$ byte-matrix,
we can finished the $4\times 4$ transpose of $M$ in one $\pshufb$ if the data $A,B,C$ and $D$
locates in the same 16-byte register.
\[
\begin{array}{c c c}
\boxed{ (a_0,a_1,a_2,a_3 ) , (b_0, b_1 ,  \ldots, d_3 ) } & \Rightarrow & 
\boxed{ (a_0,a_2,a_1,a_3) , (c_0, c_2 , \ldots  , d_3 ) } \\
\end{array} \enspace .
\]
Here the data in the same register is represented in a row box.

While the contents of matrices locate across registers, 
we can perform many transposes in parallel by swapping data between registers.
We show an example for a $4 \times 4$ transpose among 4 registers. Note that
there are $2$ swaps performed 
in each step($\Rightarrow$).
\[
\begin{array}{c c c c c}
\boxed{ (a_0,a_1, \underline{b_0, b_1} ) , \ldots } &  & 
\boxed{ (a_0,\underline{a_1}, c_0, \underline{c_1} ) , \ldots } &  & 
\boxed{ (a_0,a_2, c_0, c_2 ) , \ldots } \\ 

\boxed{ (a_2,a_3, \underline{b_2, b_3} ) , \ldots } & \Rightarrow  & 
\boxed{ (\underline{a_2},a_3, \underline{c_2}, c_3 ) , \ldots } & \Rightarrow  & 
\boxed{ (a_1,a_3, c_1, c_3 ) , \ldots } \\ 

\boxed{ ( \underline{c_0,c_1}, d_0, d_1 ) , \ldots } &  & 
\boxed{ (b_0,\underline{b_1}, d_0, \underline{d_1} ) , \ldots } &  & 
\boxed{ (b_0,b_2, d_0, d_2 ) , \ldots } \\ 

\boxed{ ( \underline{c_2,c_3}, d_2, d_3 ) , \ldots } &  & 
\boxed{ (\underline{b_2},b_3, \underline{d_2}, d_3 ) , \ldots } &  & 
\boxed{ (b_1,b_3, d_1, d_3 ) , \ldots } \\ 
\end{array} \enspace .
\]

\subsection{Bit-matrix Multiplications with Vector Instruction Set}

We also implement the matrix multiplications under AVX2 instruction set.
The matrix multiplications are performed in the $\enc$(Line~4 in Algo.~\ref{alg:encoding}) comprising
$n_p$ batches of $m\times m$ bit-matrix multiplied by $m$-bits vectors.

Algorithmically, we multiply bit-matrices with the method of the four Russians(M4R)
\cite{Aho:1974:DAC:578775}\cite{DBLP:journals/toms/AlbrechtBH10}.
Suppose we multiply a pre-defined $m \times m$ bit-matrix $\mathbf{E}$
by a $m$ bits vector $\alpha \in V_m$.
With M4R of $4$-bit, we first prepare $m/4$ tables for 
products of $\mathbf{E}$ and all vectors
in $V_4 = \mathrm{span}(v_0,\ldots,v_3)$, $\ldots$ , and $\mathrm{span}(v_{m-4},\ldots,v_{m-1})$.
By splitting $\alpha$ to $4$-bit chunks, i.e., 
$\alpha = \Sigma_{i=1}^{m/4} \alpha_i$ where $\alpha_1 \in V_4$,
$\ldots$, and $\alpha_{m/4} \in \mathrm{span}(v_{m-4},\ldots,v_{m-1})$,
we can compute $\mathbf{E} \cdot \alpha = \Sigma_{i=1}^{m/4} \mathbf{E} \cdot \alpha_i$, 
where $\mathbf{E} \cdot \alpha_i$ is obtained by one look-up of a prepared table.

In the AVX2, $\pshufb$ does exactly the look-ups for a $4$-bits indexed table.
Moreover, the $\pshufb$ performs $16$ or $32$($\vpshufb$) look-ups simultaneously.
We can detail what $\pshufb$ works in a example of $m=64$, i.e.,
multiplying $64\times 64$ bit-matrix by a $64$-bits vector.
Since each $\pshufb$ provides a $8$-bits result, 
we perform $8$ $\pshufb$ at the same $4$-bits input for the $64$-bits result,
corresponding to one $\mathbf{E}\cdot \alpha_i$.
One product of $\mathbf{E}\cdot \alpha$ comprises $m/4 = 16$ $\alpha_i$ and costs
$8\cdot 16 = 128$ $\pshufb$ in total.
Dividing by the parallelism(32) of $\vpshufb$,
one product costs $128/32=4$ $\vpshufb$ in average.

However, we have to rearrange the format the input data to work with SIMD instructions.
Given $16$ or $32$ continuing $64$-bits inputs, we collect all first bytes to first register, 
all second bytes to second register, $\ldots$ etc.
In other words, the rearrange of data is a $8\times8 $ byte-matrix transpose, 
which is also performed in the SIMD way with the method in Sec.~\ref{sec:matrix:transpose}.

\paragraph{Evaluation the effect from the point-view of data access.}

We can compare the SIMD M4R to a naive(pure memory-accessed) M4R
from the model of data access in the example of $m=64$.
The naive M4R uses $16$ memory read of $64$ bits for one product,
and costs $128$ bytes in total which is equal to SIMD M4R consuming $4 \times 32$ bytes in average.
However, the naive M4R accesses memories randomly while the SIMD M4R reads sequentially.
Our results show the SIMD M4R outperforms naive M4R.

In general, once the data of tables are read,
more parallelism in SIMD increases the performance.
However, the number of registers in CPUs restricts the parallelism to prevent 
register spilling, resulting more memory access.
In our implementation, targeting the haswell architecture in x86,
we use a 64 parallelism SIMD M4R for best performance.

%


\section{Benchmark and Discussion}
\label{sec:results}

\subsection{Benchmarks}

We benchmark our software\footnote{
The software is in \url{https://github.com/fast-crypto-lab/bitpolymul2} .}
with experiments on multiplying random bit-polynomials for 
various lengths.
Although the software is actually a constant-time implementation, i.e.,
the running time is independent of input data,
we report the average time of 100 executions.
The experiments are performed on the Intel Haswell architecture,
which is our targeting platform.
Our hardware is Intel Xeon E3-1245 v3 @3.40GHz with turbo boost disabled
and 32 GB DDR3@1600MHz memory.
The OS is ubuntu version 1604, Linux version 4.4.0-78-generic and
the compiler is gcc: 5.4.0 20160609 (Ubuntu 5.4.0-6ubuntu1~16.04.4).

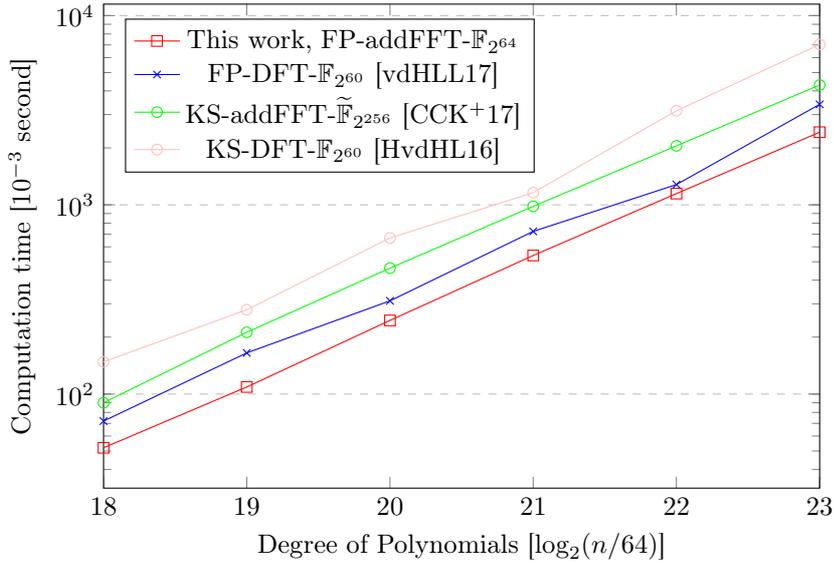
\begin{figure}[htb!]
\caption{Products in $\GF{2}[x]$ on Intel Xeon E3-1245 v3 @ 3.40GHz}
\label{fig:experiments}
\begin{tikzpicture}
\begin{semilogyaxis}[
    xlabel={Degree of Polynomials [$\log_2(n/64)$]},
    ylabel={Computation time [$10^{-3}$ second]},
    xmin=18, xmax=23,
    xtick={18,19,20,21,22,23},
    legend pos=north west,
    ymajorgrids=true,
    grid style=dashed,
    height=8cm,
    width=11cm,
]
 
\addplot[
    color=red,
    mark=square,
    ]
    coordinates {
    (18,52)(19,109)(20,245)(21,540)(22,1147)(23,2420)
    };
    \addlegendentry{This work, FP-addFFT-$\GF{2^{64}}$}
\addplot[
    color=blue,
    mark=x,
    ]
    coordinates {
    (18,72)(19,165)(20,311)(21,724)(22,1280)(23,3397)
    };
    \addlegendentry{FP-DFT-$\GF{2^{60}}$ \cite{VLL2017:carrylesspolymul}}
\addplot[
    color=green,
    mark=o,
    ]
    coordinates {
    (18,90)(19,212)(20,463)(21,982)(22,2050)(23,4299) 
    };
    \addlegendentry{KS-addFFT-$\TGF{2^{256}}$ \cite{DBLP:journals/corr/abs-1708-09746}}
\addplot[
    color=pink,
    mark=o,
    ]
    coordinates {
    (18,148)(19,279)(20,668)(21,1160)(22,3142)(23,7040) 
    };
    \addlegendentry{KS-DFT-$\GF{2^{60}}$ \cite{DBLP:conf/issac/HarveyHL16}}
\end{semilogyaxis}
\end{tikzpicture}
\end{figure}

\begin{table}[htb!] 
\begin{center}
\begin{threeparttable}
\caption{ 
Products in $\GF{2}[x]$ on Intel Xeon E3-1245 v3 @ 3.40GHz ($10^{-3}$ sec.)
The implementations in upper table base on Frobenius partitions
and the lower implementations are with Kronecker substitution.
}
\label{tab:bitpolymul}
\begin{tabular}{|l | r | r| r | r | r | r | r | r | }
\hline
   $\log_2 n/64$ & 16  & 17 &  18 & 19 & 20 & 21 & 22 & 23   \\  \hline\hline
This work, $\GF{2^{64}}$ \tnote{a} 
& 12 & 25 & 52 & 109 & 245 & 540 & 1147 & 2420 \\  \hline
This work, $\GF{2^{128}}$ \tnote{a} 
& 13 & 28 & 58 & 123 & 273 & 589 & 1248 & 2641 \\  \hline
DFT, $\GF{2^{60}}$ \cite{VLL2017:carrylesspolymul} \tnote{b} 
&  15 &  32  &  72   & 165   & 311   &  724  & 1280 & 3397 \\  \hline
\hline
\hline
KS-$\TGF{2^{256}}$ \cite{DBLP:journals/corr/abs-1708-09746} \tnote{c}    
& 20 & 41 &  93 & 216 & 465 & 987 & 2054 & 4297  \\ \hline 
KS-$\GF{2^{128}}$\cite{DBLP:journals/corr/abs-1708-09746} \tnote{c}  
& 25 & 53 & 115 & 252 & 533 & 1147 & 2415 & 5115 \\ \hline
KS-$\GF{2^{60}}$ \cite{DBLP:conf/issac/HarveyHL16} \tnote{d} 
&  29 &  64  &  148   & 279   & 668   &  1160  & 3142 & 7040 \\ \hline
\texttt{gf2x} \cite{gf2x} \tnote{e}  
& 26 & 59 & 123 & 285 & 586 & 1371 & 3653 & 7364  \\ \hline 
\hline
\end{tabular}
\begin{tablenotes}
  \small
  \item [a] Version 1656d5e. \url{https://github.com/fast-crypto-lab/bitpolymul2}
  \item [b] SVN r10681. Available from \url{svn://scm.gforge.inria.fr/svn/mmx}
  \item [c] Version c13769d. \url{https://github.com/fast-crypto-lab/bitpolymul}
  \item [d] SVN r10663. Available from \url{svn://scm.gforge.inria.fr/svn/mmx}
  \item [e] Version 1.2.  Available from \url{http://gf2x.gforge.inria.fr/}
\end{tablenotes}
\end{threeparttable}
\end{center}
\end{table}
%


Figure~\ref{fig:experiments} shows the results of our experiments
and the comparisons with previous implementations.
The figure shows the running time verse degree of polynomials both
in logarithm scales.
The ``FP'' and ``KS'' stands for Frobenius partition and Kronecker substitution respectively.
More details about the results can be found in Tab.~\ref{tab:bitpolymul}.

The result shows that our implementations clearly outperform all previous implementations.
Among our implementations, the version of $\GF{2^{64}}$ is faster than $\GF{2^{128}}$
for more efficient multiplications over underlying fields.
However, the version of $\GF{2^{128}}$ supports polynomials of larger degree.
From the figure, we can see the same tendency among all data.
This suggests that these algorithms work roughly in the same
complexity level while our implementation, however, works with lowest hidden constant.
We can also see the straight lines for additive FFT based algorithms, but the
line turns slightly for the multiplicative algorithms.
It is caused from that polynomials with terms of 2 powers are not optimal for
particular sizes of multiplicative groups.
Lastly, from the values in the table, we can see the FP implementations
lead KS implementations about the factor of two, 
which is consistent with the conclusion of \cite{VLL2017:carrylesspolymul}.


\subsection{Profiles}

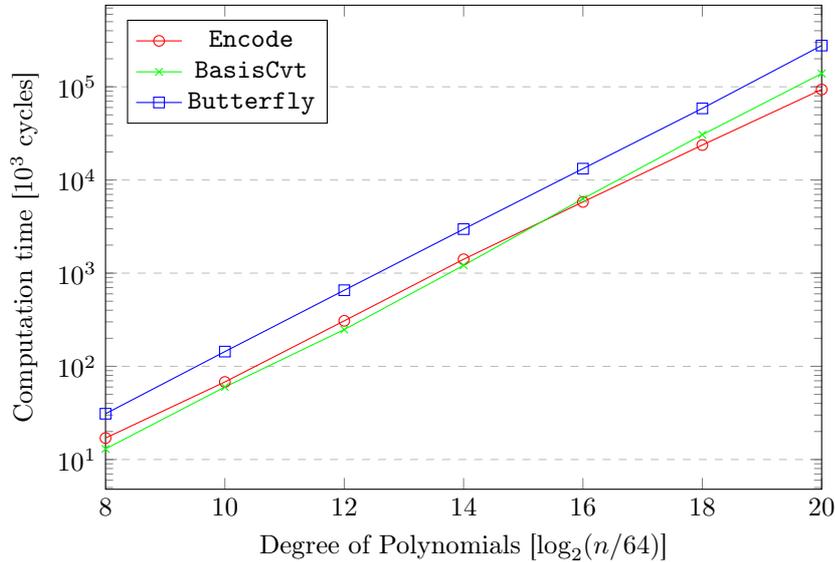
\begin{figure}[htb!]
\caption{Profiles for multiplications in $\GF{2}[x]$ on Intel Xeon E3-1245 v3 @ 3.40GHz}
\label{fig:profiles}
\begin{tikzpicture}
\begin{semilogyaxis}[
    xlabel={Degree of Polynomials [$\log_2(n/64)$]},
    ylabel={Computation time [$10^{3}$ cycles]},
    xmin=8, xmax=20,
    xtick={ 8 , 10 , 12 , 14 , 16 , 18 , 20 },
    legend pos=north west,
    ymajorgrids=true,
    grid style=dashed,
    height=8cm,
    width=11cm,
]
 
\addplot[
    color=red,
    mark=o,
    ]
    coordinates {
    (8,17)(10,68)(12,308)(14,1410)(16,5816)(18,23720)(20,93572)
    };
    \addlegendentry{$\enc$}
\addplot[
    color=green,
    mark=x,
    ]
    coordinates {
    (8,13)(10,60)(12,248)(14,1212)(16,6292)(18,30619)(20,138829)
    };
    \addlegendentry{$\basiscvt$}
\addplot[
    color=blue,
    mark=square,
    ]
    coordinates {
    (8,31)(10,144)(12,656)(14,2971)(16,13237)(18,58706)(20,277221)
    };
    \addlegendentry{$\lchbtfy$}
\end{semilogyaxis}
\end{tikzpicture}
\end{figure}

We show the profiles of algo.~\ref{alg:polymul} for $\encode$, $\basiscvt$, and $\lchbtfy$
in Tab.~\ref{tab:profiles} and Fig.~\ref{fig:profiles}
since the 3 components actually work in different levels of complexities.
The complexities, ordered by levels, are $O(n \log n \log \log n)$, $O(n \log n)$, and $O(n)$
 for $\basiscvt$, $\lchbtfy$, and $\encode$ respectively. 
In the practical range of polynomial lengths, however,
the $\lchbtfy$ costs the most computation time although it is not
the dominant term of the complexities.
We can also see the running time of $\basiscvt$ does increase faster than $\encode$
from the Fig.~\ref{fig:profiles}.


\begin{table}[h] 
\begin{center}
\begin{threeparttable}
\caption{ 
Profiles for multiplications in $\GF{2}[x]$ on Intel Xeon E3-1245 v3 @ 3.40GHz ($10^3$ cycles).}
\label{tab:profiles}
\begin{tabular}{|l | r | r| r | r | r | r | r | }
\hline
   $\log_2 n/64$ & 8  & 10 &  12 & 14 & 16 & 18 & 20   \\  \hline\hline
$\enc $ 
& 17 & 68 & 308 & 1410 & 5816 & 23720 & 93572 \\  \hline
$\basiscvt$ 
& 13 & 60 & 248 & 1212 & 6292 & 30619 & 138829 \\  \hline
$\lchbtfy$
& 31 &  144  & 656 & 2971 & 13237 & 58706 & 277221 \\  \hline
\end{tabular}
\end{threeparttable}
\end{center}
\end{table}


\section{Summary}

We have shown the new algorithm for multiplying bit-polynomials of large degrees
as well as its implementation with SIMD instructions.
The new algorithm is based on evaluating polynomials at
the Frobenius partition $\Sigma = v_{l+m/2} + V_l$ with the additive FFT.
This form of partition particularly fits the additive FFT.
A new process $\enc$ accelerates the $\lchbtfy$ by performing
the $l_m$ layers of butterflies as matrix multiplications
and truncating the unnecessary results for further $\lchbtfy$.

For implementing the algorithm, we show the efficient memory access models and
the SIMD implementation of the
key components(e.g., bit-matrix transpose and bit-matrix multiplication).
The multiplications over underlying fields are also designed to utilize
the $\pclmul$ instruction.
At last, the experiments show our software outperforms all previous implementations
to the best of our knowledge.





\newcommand{\etalchar}[1]{$^{#1}$}
\providecommand{\skiptext}[1]{} \hyphenation{ASIA-CRYPT}

\appendix

\end{document}